\documentclass[aps,prc,twocolumn,showpacs,preprintnumbers,amsmath,amssymb,superscriptaddress]{revtex4-1}

\usepackage{graphicx}
\usepackage{epsfig}
\usepackage{dcolumn}
\usepackage{bm}
\usepackage{epstopdf}

\begin{document}


\title
{Nature of low-lying electric dipole resonance excitations in $^{74}$Ge}

\author{D.~Negi} 
\email{dinphysics@gmail.com} 
\affiliation{iThemba LABS, P.O. Box 722, Somerset West 7129, South Africa}
\affiliation{UM-DAE Centre for Excellence in Basic Sciences, Mumbai 400098, India }
\author{ M.~Wiedeking} 
\email{wiedeking@tlabs.ac.za}
\affiliation{iThemba LABS, P.O. Box 722, Somerset West 7129, South Africa}
\author{E.~G.~Lanza}
\affiliation{INFN, Sezione di Catania, I-95123 Catania, Italy}
\author {E.~Litvinova}
\affiliation{Western Michigan University, Kalamazoo, MI 49008-5252, USA}
\affiliation{National Superconducting Cyclotron Laboratory, Michigan State University, East Lansing, Michigan 48824-1321, USA}
\author{ A.~Vitturi}
\affiliation{Dipartimento di Fisica e Astronomia, Universit{\`a} di Padova, Italy}
\affiliation{INFN, Sezione di Padova, I-35131 Padova, Italy}
\author{ R.~A.~Bark}
\affiliation{iThemba LABS, P.O. Box 722, Somerset West 7129, South Africa}
\author{ L.~A.~Bernstein}
\affiliation{Lawrence Livermore National Laboratory, Livermore, CA 94550-9234, USA}
\affiliation{University of California, Berkeley, CA 94720-1730, USA}
\author{D.~L.~Bleuel} 
\affiliation{Lawrence Livermore National Laboratory, Livermore, CA 94550-9234, USA}
\author{S.~Bvumbi}
\affiliation{University of Johannesburg, Auckland Park 2006, South Africa}
\author{T.~D.~Bucher}
\affiliation{iThemba LABS, P.O. Box 722, Somerset West 7129, South Africa}
\author{ B.~H.~Daub}
\affiliation{Lawrence Livermore National Laboratory, Livermore, CA 94550-9234, USA}
\affiliation{University of California, Berkeley, CA 94720-1730, USA}
\author{T.~S.~Dinoko}
\affiliation{iThemba LABS, P.O. Box 722, Somerset West 7129, South Africa}
\affiliation{University of the Western Cape, Bellville 7535, South Africa }
\author{J.~L.~Easton}
\affiliation{iThemba LABS, P.O. Box 722, Somerset West 7129, South Africa}
\affiliation{University of the Western Cape, Bellville 7535, South Africa }
 \author{A.~G{\"o}rgen} 
\affiliation{Department of Physics, University of Oslo, N-0316 Oslo, Norway}
\author{M.~Guttormsen}
\affiliation{Department of Physics, University of Oslo, N-0316 Oslo, Norway}
\author{ P.~Jones}
\affiliation{iThemba LABS, P.O. Box 722, Somerset West 7129, South Africa}
\author{ B.~V.~Kheswa}
\affiliation{iThemba LABS, P.O. Box 722, Somerset West 7129, South Africa}
\affiliation{Department of Physics, Stellenbosch University, Matieland 7602, South Africa}
\author{ N.~A.~Khumalo} 
\affiliation{University of the Western Cape, Bellville 7535, South Africa }
\affiliation{University of Zululand, KwaDlangezwa 3886, South Africa}
\author{A.~C.~Larsen} 
\affiliation{Department of Physics, University of Oslo, N-0316 Oslo, Norway}
\author{E.~A.~Lawrie}
\affiliation{iThemba LABS, P.O. Box 722, Somerset West 7129, South Africa}
\author{ J.~J.~Lawrie}
\affiliation{iThemba LABS, P.O. Box 722, Somerset West 7129, South Africa}
\author{S.~N.~T.~Majola} 
\affiliation{iThemba LABS, P.O. Box 722, Somerset West 7129, South Africa}
\affiliation{University of Cape Town, Rondebosch 7701, South Africa}
\author{L.~P.~Masiteng}
\affiliation{University of Johannesburg, Auckland Park 2006, South Africa}
\author{M.~R.~Nchodu} 
\affiliation{iThemba LABS, P.O. Box 722, Somerset West 7129, South Africa}
\author{J.~Ndayishimye}
\affiliation{iThemba LABS, P.O. Box 722, Somerset West 7129, South Africa} 
\affiliation{Department of Physics, Stellenbosch University, Matieland 7602, South Africa}
\author{R.~T.~Newman}
\affiliation{Department of Physics, Stellenbosch University, Matieland 7602, South Africa}
\author{S.~P.~Noncolela} 
\affiliation{iThemba LABS, P.O. Box 722, Somerset West 7129, South Africa}
\affiliation{University of the Western Cape, Bellville 7535, South Africa }
\author{J.~N.~Orce} 
\affiliation{University of the Western Cape, Bellville 7535, South Africa }
\author{P.~Papka}
\affiliation{iThemba LABS, P.O. Box 722, Somerset West 7129, South Africa}
\affiliation{Department of Physics, Stellenbosch University, Matieland 7602, South Africa}
\author{L.~Pellegri}
\affiliation{iThemba LABS, P.O. Box 722, Somerset West 7129, South Africa} 
\affiliation{University of the Witwatersrand, Johannesburg 2050, South Africa}
\author{ T.~Renstr{\o}m}
\affiliation{Department of Physics, University of Oslo, N-0316 Oslo, Norway}
\author{ D.~G.~Roux}
\affiliation{Rhodes University, Grahamstown 6410, South Africa}
\author{R.~Schwengner}
\affiliation{Helmholtz-Zentrum Dresden-Rossendorf, 01328 Dresden, Germany}
\author{O.~Shirinda} 
\affiliation{iThemba LABS, P.O. Box 722, Somerset West 7129, South Africa}
\affiliation{Department of Physics, Stellenbosch University, Matieland 7602, South Africa}
\author{S.~Siem}
\affiliation{Department of Physics, University of Oslo, N-0316 Oslo, Norway}

\date{\today}

\begin{abstract}

Isospin properties of dipole excitations in $^{74}$Ge are investigated using the ($\alpha$, $\alpha'$$\gamma$) 
reaction and compared to ($\gamma$, $\gamma '$) data. The results indicate that the dipole 
excitations in the energy region of 6 to 9 MeV adhere to the scenario of the recently found splitting of the 
region of dipole excitations into two separated parts: one at low energy being populated by both isoscalar and 
isovector probes and the other at high energy, excited only by the electromagnetic probe. RQTBA calculations
show a reduction in the isoscalar $E1$ strength with an increase in excitation energy which is consistent with the measurement.

\end{abstract}


\maketitle

\section{\label{sec:level1}INTRODUCTION}

In recent years there has been a surge in experimental studies of dipole excitations
lying on the low-energy tail of the isovector giant dipole resonance, the so-called Pygmy dipole
resonance (PDR). The PDR has been interpreted as an exotic mode of excitation  
due to the motion of a weakly bound neutron excess against an almost inert proton-neutron core \cite{paar2007,savran2013, bracco2015}, 
although single particle-hole excitations are also considered \cite{lane1971,reinhard2013}.
One major reason for the renewed interest in the PDR is the possibility of carrying out high-resolution
measurements on these low-lying dipole excitations using heavy ion \cite{pellegri2014,crespi2015}, proton \cite{poltoratska2012,krumbholz2015}, and $\alpha$ inelastic scattering experiments \cite{savran2006,endres2009}.
An experimental technique, combining particle and $\gamma$-ray detection techniques, to study the response of dipole excitations to isoscalar probes
was pioneered by Poelhekken \textit{et al.} \cite{poelhekken1992} and applied in several studies
since \cite{savran2006,endres2009,endres2010,derya2012,derya2013,derya2014,crespi2014,pellegri2014,crespi2015}.
These experiments provide complementary information to those obtained from ($\gamma$, $\gamma '$) experiments which investigate the isovector nature of the
excitations \cite{volz2006,govaert1998,romig2013,shizuma2008,schw2007,ben2009,mak2010}. One of the surprising results
from recent experiments is the isospin splitting of the PDR \cite{savran2013,bracco2015,savran2006,endres2009,endres2010}. This provides intimate knowledge
about the isospin nature of these excitations which would not be possible to infer from ($\gamma$, $\gamma '$) experiments alone. These experimental discoveries were followed by intensive theoretical
investigations \cite{tsoneva2008,paar2009,martini2011,tohyama2012,vretenar2012,nakada2013}. 

Incidentally, scattering experiments with isocalar probes for the study of the PDR have so far been limited to
only certain regions of the nuclear chart and carried out mainly on nuclei with large neutron-to-proton ratios \cite{savran2006,endres2010,derya2012,crespi2014, pellegri2014}.
Information on how the results from scattering reactions compare to those of
($\gamma$, $\gamma '$) experiments in nuclei closer to $N/Z =$1 are also becoming available \cite{poelhekken1992,derya2013,derya2014,crespi2015}. Since most of the incident
isoscalar probes are sensitive to the surface of the nucleus, the information gathered advances our understanding of the evolution of the PDR with changing $N/Z$. This information is extrapolated for obtaining better estimates of the total strength
exhausted by the PDR in nuclei of astrophysical importance, many of which are still inaccessible with the 
available experimental facilities and techniques. The PDR has been suggested to have a significant
impact on neutron capture rates and isotopic solar abundance distributions in r-process
nucleosynthesis \cite{goriely1998,goriely2004,litvinova2009,daoutidis2012}. Further, the PDR could possibly constrain
the equation-of-state of hot and dense neutron matter as found in neutron star remnants \cite{piekarewicz2006,piekarewicz2011}. 

In this contribution, we present results on $^{74}$Ge where a high-resolution measurement was carried out using the $\alpha$ inelastic scattering reaction. In its ground state, $^{74}$Ge is a
moderately deformed prolate nucleus \cite{rosier1989,moeller2008} with $N/Z=$ 1.32. For comparison and to facilitate
the discussion, information about the $E1$ strength distribution is also available from ($\gamma$, $\gamma '$) data in $^{74}$Ge \cite{jung1995,massarczyk2015}.

\section{EXPERIMENTAL DETAILS}

The experiment was performed at the Separated Sector Cyclotron facility at iThemba LABS with the
AFRican Omnipurpose Detector for Innovative Techniques and Experiments (AFRODITE)
$\gamma$-ray detector array \cite{afrodite} in conjunction with two identical
particle-telescopes, each of them consisting of two silicon detectors (in $\Delta E-E$
configuration). The $\alpha-$particles with a beam energy
of 48 MeV impinged on a 500 $\mu$g/cm$^2$ thick $^{74}$Ge target to populate excited states in the
inelastic scattering reaction. The experiment was carried out over a period of five days with an
average beam current of $\sim$ 14 particle nA. The telescopes were placed at an angle of $\theta = \pm$45$^{\circ}$
with respect 
to the beam axis. The dimensions of the W1-type double sided silicon strip detectors \cite{micron} were 5 cm $\times$
5 cm and consisted of 16 parallel and perpendicular strips 3 mm wide. The distance from target to the
telescopes was 5 cm yielding an angular range of $20^\circ$ to $72^\circ$
in the laboratory frame of reference. Thicknesses of the $\Delta$\textit{E} and \textit{E} detectors were
284 and 1000
$\mu$m, respectively, and to suppress $\delta$ electrons an aluminum foil of 4.1 mg/cm$^2$ areal density 
was placed in front of
the $\Delta$\textit{E} detectors. Calibration of individual strips of the silicon detectors was
performed using a $^{228}$Th $\alpha$ source.

AFRODITE, at the time of the experiment, consisted of nine Clover HPGe detectors with four detectors at
135$^\circ$ and five
at 90$^\circ$ at a distance of 19.6 cm from the target. The detectors were calibrated using standard
$^{152}$Eu and $^{56}$Co sources.  High $\gamma$-ray energy efficiency parameters for the AFRODITE
array were available from Ref.~\cite{afrodite}. XIA digital electronics \cite{XIA} was used to acquire
timestamped online data in singles mode.

\section{DATA ANALYSIS}

\begin{figure}[t]
\includegraphics[scale=0.4]{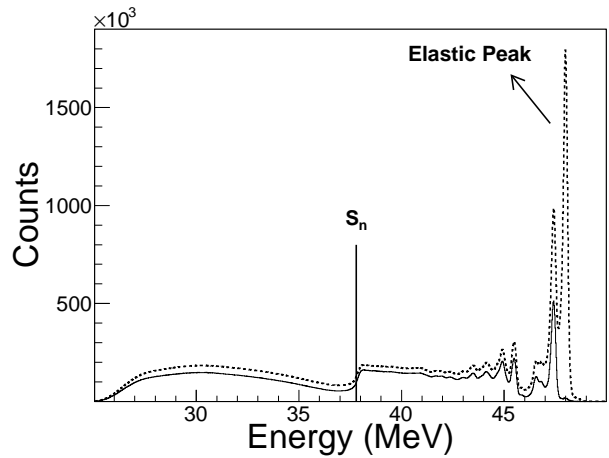}
\caption{\label{fig:Particle-proj} Spectrum of $\alpha$ particles detected in coincidence
with $\gamma-$rays. Solid and dashed curves are representing  data with and without the subtraction of
uncorrelated events, respectively. Visible peaks (solid curve) are strongly populated discrete states in $^{74}$Ge. S$_n$ indicates the location of the neutron separation energy. }
\end{figure}

From the timestamped data, events with single, double, and higher fold coincidences were
constructed with an offline coincidence time window of 600 ns. From double fold events, the $\alpha-\gamma$ coincidences
were extracted by placing a gate on the $\alpha-$particles in the particle identification spectrum.
A projection of $\alpha-\gamma$ coincidences onto the $\alpha-$particle axis is shown in Fig.~\ref{fig:Particle-proj}.
The selection of correlated events was made with a coincidence time of less than 140 ns by placing
appropriate gates around the prompt time peak. Uncorrelated event contributions were extracted and
subtracted from the data by placing off-prompt time gates to the early and late sides of the prompt timing
peak. 

Kinematic corrections due to the recoil energy of $^{74}$Ge and the energy losses of
scattered $\alpha-$particles in the target and aluminum foils were applied to the $\alpha-$particles. Although the target contained some oxygen and carbon contaminants, the
recoil corrections for the scattered $\alpha-$particles from $^{74}$Ge are quite different compared to
those of light contaminant nuclei, thereby allowing a clean extraction of the events of interest. For
instance, the corrections from $^{74}$Ge versus $^{16}$O differ by $\sim$ 1 MeV 
and $\sim$ 10 MeV at $20^\circ$ and $72^\circ$ detection angles, respectively. The energy resolution of
the $\Delta E-E$ telescopes, measured from the elastic peak, was $\approx$ 250 keV. Despite the low
velocities of the $^{74}$Ge recoils, corrections for Doppler effects of the high-energy $\gamma-$rays were found to be necessary and useful. 

Transitions ($E_{\gamma}$) to the ground state were extracted with the condition $|E_{\gamma}-E_x | \leq $ 130 keV imposed on the
$\alpha-\gamma$ coincidence events, where  $E_x$ refers to the excitation energy of the decaying state
and is determined from the energy of the scattered $\alpha-$particles.
Placing this stringent energy requirement upon the data, together with the differences in kinematic
properties ensures that only transitions from $^{74}$Ge are extracted, eliminating contributions due to contaminants. 

Additionally, various combination of angles between the direction of the recoiling nuclei
(as defined by the $\alpha$ particles detected in the particle telescope) and the
$\gamma$-rays detected in the Clover detectors were used for the determination of angular distributions.

\section{RESULTS AND DISCUSSION}

\begin{figure}[t]
\includegraphics[scale=0.4]{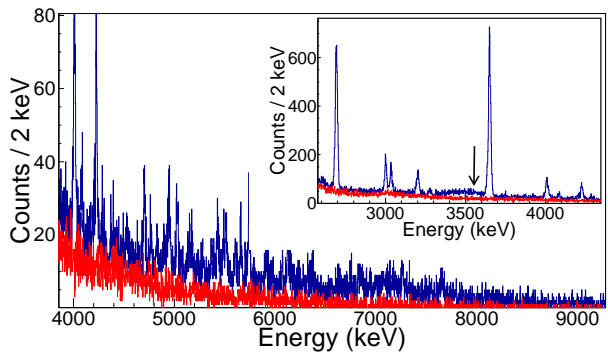}
\caption{\label{fig:Gamma-spectrum} (Color online) Spectrum of $\gamma$-ray transitions decaying directly
to the ground state from defined excitation energies. Blue and red spectra correspond to correlated
and uncorrelated events, respectively.
Inset: the lower energy part of the spectrum where the arrow indicates the position of the unobserved
3558-keV transition, known from ($\gamma$, $\gamma '$) experiments \cite{jung1995,massarczyk2015}.}
\end{figure}

The spectrum of direct $\gamma$-ray transitions to the ground state is shown in Fig.~\ref{fig:Gamma-spectrum}, where in addition to many states for $E_x <$ 6 MeV,
a high concentration of states/strength is also observed for 6.5 $< E_x <$ 8 MeV. Although the overall sensitivity to
high-energy transitions is relatively poor, many transitions observed in ($\gamma$, $\gamma '$) experiments
\cite{jung1995,massarczyk2015} can also be clearly identified in the 
present data. Unresolved strength was separated from intensities of individual transitions by simultaneously fitting the peaks using
the ROOT analysis package \cite{root} in a 16 keV per channel compressed $\gamma-$ray spectrum. The
unresolved, underlying intensity for 6.5 $< E_x <$ 8 MeV amounts to $\approx 50 \%$.
Comparisons with the recent ($\gamma$, $\gamma '$) measurement \cite{massarczyk2015} reveal 
several states, which were not populated in the ($\alpha$, $\alpha'$$\gamma$) reaction, but are observed in the ($\gamma$, $\gamma '$) measurement. However, the states at
$E_x =$ 6850 and 7060 keV are populated only through the ($\alpha$, $\alpha'$$\gamma$) reaction.

\begin{figure}[t]
\includegraphics[scale=0.5]{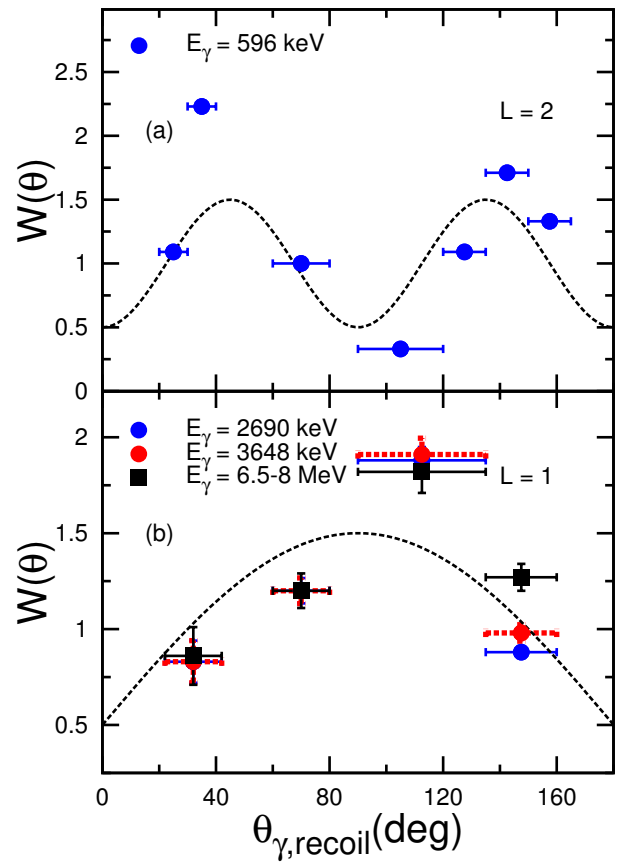}
\caption{\label{fig:ang_dist} (Color online) Angular distributions of (a) the first-excited state
 $l=2$ 596-keV transition, and (b) $l=1$ transitions from the known
 2690 and 3648 keV states together with the total strength of resolved and unresolved transitions for 6.5 $< E_x <$ 8 MeV
 in $^{74}$Ge.}
\end{figure}

The multipole nature of the high-energy transitions was determined through angular distribution
measurements, shown in Fig.~\ref{fig:ang_dist}. Because of the paucity of the data, the angular
distribution was extracted simultaneously for the total (resolved and unresolved) $\gamma$-ray strength in the interval 6.5 $< E_x <$ 8 MeV.
For comparison, angular distributions of known dipole ($E_{\gamma}$ = 2690 keV and $E_{\gamma}$ = 3648 keV) and quadrupole
transitions ($E_{\gamma}$ = 596 keV) in $^{74}$Ge are also included in Fig.~\ref{fig:ang_dist}.
Although the 6.5 $< E_x <$ 8 MeV strength does not exhibit a perfect agreement with the expected distribution of a dipole transition,
the similarity to the two known dipole transitions strongly supports the overall dipole nature. Natural-parity states are  preferentially
populated in this reaction \cite{eidson1962}, leading to an assignment of spin-parity $J^{\pi}=1^-$ to the decaying states.

\begin{figure}[t] 
\includegraphics[scale=0.3]{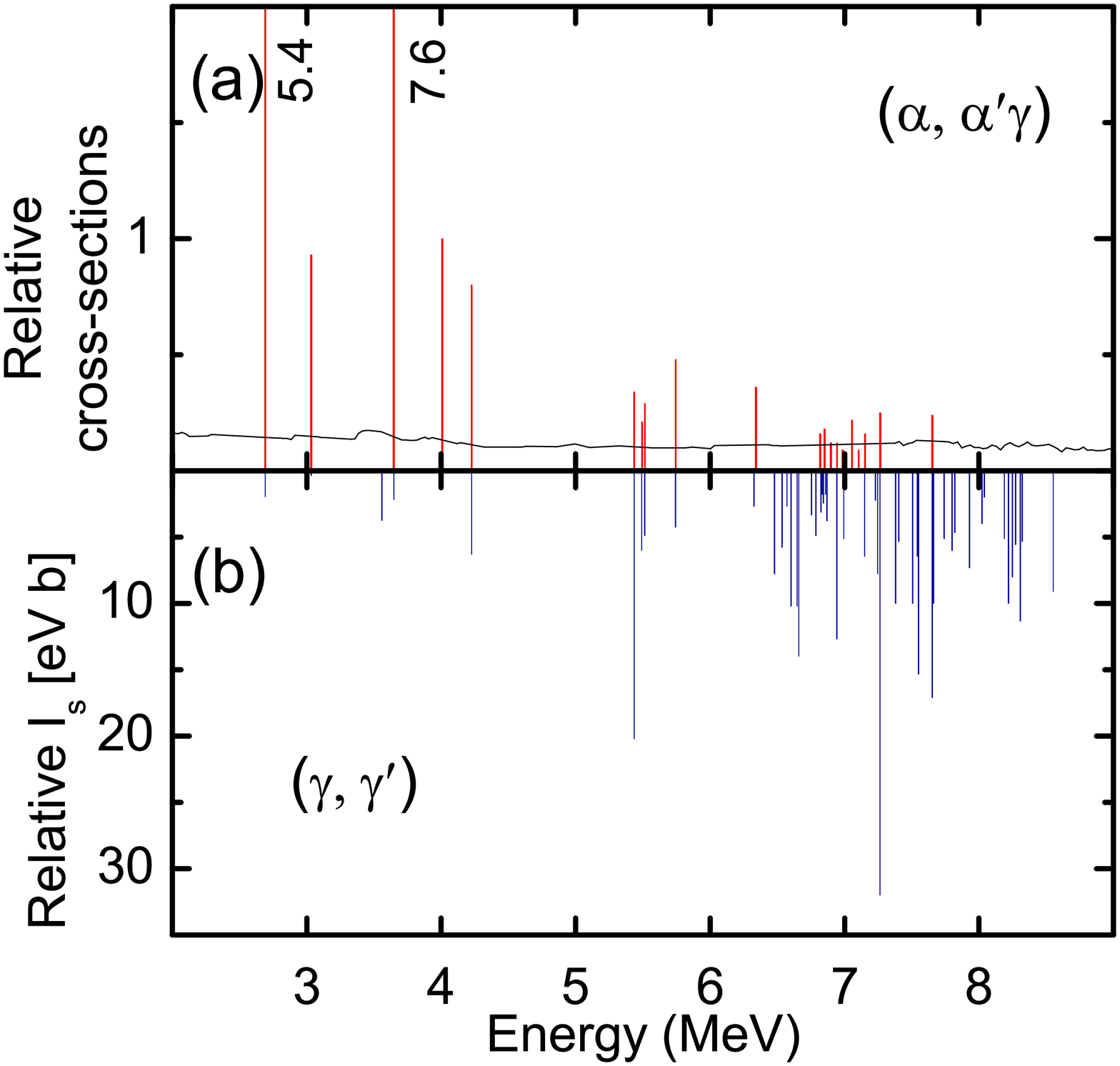}
\caption{\label{fig:alpha_compare} (Color online) In panel (a) relative cross-sections of $E1$ transitions
from the ($\alpha$, $\alpha'$$\gamma$) data are plotted, while in panel (b) the relative integrated
scattering cross-sections $I_s$ obtained
from ($\gamma$, $\gamma '$) data \cite{massarczyk2015} are shown. Numbers next to some transitions indicate the total value of relative cross section.
In panel (a), the sensitivity limit is shown by the black solid curve and was determined using the procedure outlined in Ref.~\cite{endres2009}.
Uncertainties on the cross-sections in panel (a) are $\sim50\%$ for weakly populated states and decrease to $\sim15\%$ for strongly populated states.}
\end{figure}

In Fig.~\ref{fig:alpha_compare} panel (a), relative cross-sections of observed  $J^\pi = 1^-$ states are plotted and normalized to the 4007 keV state. 
For comparison, panel (b) of Fig.~\ref{fig:alpha_compare} displays relative integrated scattering cross-sections ($I_s$)
from ($\gamma$, $\gamma '$) data  \cite{massarczyk2015}, where the 4007 keV state is taken as the 
reference once again. All states for $E_x >$ 6 MeV from the ($\gamma$, $\gamma '$) data are assumed to have negative parity and are plotted in Fig.~\ref{fig:alpha_compare}, whereas in both panels only states have been included
with known negative parity for $E_x <$ 6 MeV, as deduced from the ($\gamma$, $\gamma '$) data. An exception are the states 2690, 3033, and 3648 keV with assigned $J^{\pi} =1, 1, 1^+$, respectively \cite{jung1995}. The $J^{\pi} =1^+$ assignment to the 3648 keV state is based
on a polarization measurement \cite{jung1995}. However, this state has also been observed in an earlier ($\alpha$, $\alpha'$) work \cite{schurmann1987}.
Since inelastic $\alpha$-scattering populates preferentially natural-parity states, the observed strong cross section in the present experiment contradicts this assignment.
Hence, the transition is assumed to be electric dipole in character. 
Similar considerations are applied to the 2690 and 3033 keV states. The complete
absence of the 3558 keV state in the present data (see arrow in inset of Fig.~\ref{fig:Gamma-spectrum}) is noteworthy,
since this state has been observed in the ($\gamma$, $\gamma '$) work and was assigned
$J^\pi = 1^{(-)}$ ~\cite{jung1995}.

The comparison shows the presence of two different regions in the energy range of the investigated dipole excitations. In the lower part (3 $<E_x <$ 6 MeV) the excitations
due to ($\alpha$, $\alpha'$$\gamma$) are enhanced compared to the upper part (6 $<E_x <$ 9 MeV). For
($\gamma$, $\gamma '$) excitations the trend is reversed indicating a dominant isovector nature of the higher-energy
dipole excitations. This reduction in relative cross-section in the ($\alpha$, $\alpha'$$\gamma$) data becomes even more
pronounced if the intensity of the 3648-keV state is
taken as a normalization reference.

The reduction of cross-sections in the ($\alpha$, $\alpha'$$\gamma$) data for states $E_x>$ 6 MeV, compared to cross-sections for $E_x<$ 6 MeV,
is larger than observed in previous cases. Indeed,
with respect to ($\alpha$, $\alpha'$$\gamma$) studies on $^{140}$Ce, $^{138}$Ba and $^{124}$Sn \cite{savran2006,endres2009,endres2010},
the isoscalar response at low energies ($<$ 6 MeV) is much stronger. The current result shows that many of the 
dipole excitations in the 6 $< E_x <$ 9 MeV range in $^{74}$Ge are mixed with larger isovector components.
However, a few weakly populated pure isoscalar, as well as several pure isovector states are found for $E_x > 6$ MeV.
These results indicate that the dipole excitations in $^{74}$Ge for $E_x>$ 6 MeV do show the common scenario
of dipole excitations splitting in two distinct parts: one at lower energy, whose states have a strong isospin mixing, and one
at higher energy with predominant isovector character. 

We have performed calculations of the dipole transition densities in $^{74}$Ge 
within the Relativistic Quasiparticle Time Blocking Approximation (RQTBA) \cite{litvinova2008} based
on the covariant energy density functional theory (CEDFT) \cite{ring1996,vretenar2005}.
The RQTBA has been developed to include spreading mechanisms, other
than Landau damping (one particle - one hole (1p1h), or two-quasiparticle (2q), configurations)
into the microscopic description of nuclear excitation modes within the relativistic framework.
The existing versions of RQTBA include 2q $\otimes$ phonon \cite{litvinova2008} or two phonon \cite{litvinova2010,litvinova2013}
configurations in a fully self-consistent way. Parameters (in the present version with the
NL3$^*$ \cite{lalazissis2009} interaction - 8 parameters) of the CEDFT were fixed by fitting
masses and radii of several characteristic nuclei throughout the nuclear chart \cite{vretenar2005} 
and no adjustments were involved in the subsequent calculations. 

\begin{figure}[t]
\includegraphics[scale=0.45]{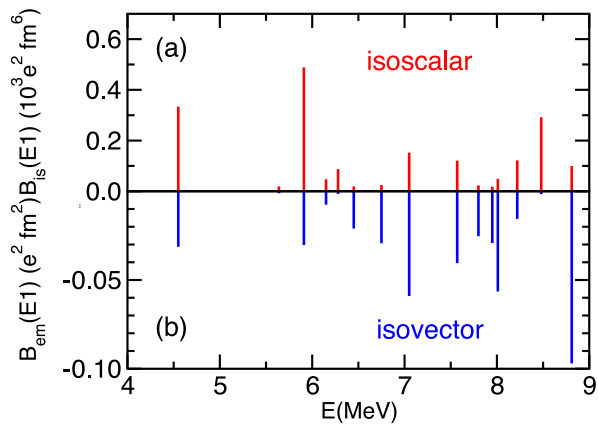}
\caption{\label{fig:calculations_compare} (Color online) Reduced transition probabilities in $^{74}$Ge 
from RQTBA calculations plotted for the isoscalar (a) and electromagnetic (isovector) (b) dipole operators.}
\end{figure}

The calculations were performed
in the following three steps: (i) the single-particle spectrum was obtained from the self-consistent
relativistic mean-field solution; (ii) the phonon spectrum was computed by the self-consistent
relativistic quasiparticle random phase approximation (RQRPA) and (iii) the Bethe-Salpeter equation
for the nuclear dipole response was solved within the RQTBA employing the RQRPA phonons
to construct the induced energy-dependent residual interaction. The low-energy region of the dipole spectrum is calculated with the RQTBA. It includes mixing of
quasiparticles with phonons, in particular, with the lowest 2$^+$ collective state obtained in RQRPA
at $E_x \sim$0.6 MeV and the lowest 3$^-$ state at $E_x \sim$3.4 MeV, while without mixing there is no dipole strength
at the energies of interest. The phonon spectra are consistent with experimental observations for the first-excited 2$^+$ and 3$^-$ states at 596 and 2536 keV \cite{nndc}.  Reduced transition probabilities from RQTBA
calculations with 25 keV smearing (bunching) for isoscalar and isovector dipole operators are
plotted in panels (a) and (b) of Fig.~\ref{fig:calculations_compare}. Although these calculations also
suggest a suppression in the isoscalar $E1$ strength at higher
energies, they underestimate the experimentally observed suppression in $^{74}$Ge.
 
\begin{figure}[b]
\includegraphics[scale=0.35]{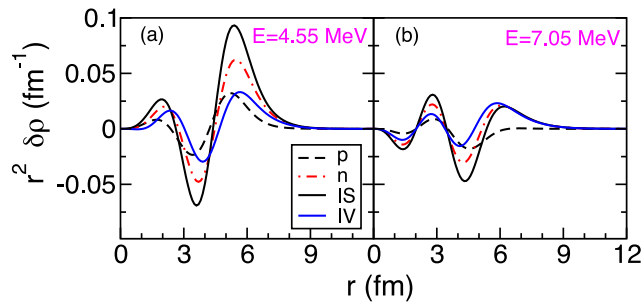}
\caption{\label{fig:td} (Color online) Transition densities for two calculated RQTBA states  at $E_x$=4.55 (panel (a)) and 7.05 MeV (panel(b))  in $^{74}$Ge.}
\end{figure}

Figure~\ref{fig:td} shows the proton, neutron, isoscalar and isovector transition densities for calculated states
at $E_x =$ 4.55 and 7.05 MeV. The lower-lying state
(panel (a)) exhibits the usual pattern for an almost pure isoscalar dipole state, with the proton and
neutron transition densities in phase inside
the nucleus and at the nuclear surface. Consequently, the isoscalar transition density has a pattern
typical of the compressional mode with a node close to the nuclear surface. In contrast, the higher-lying state (panel (b)) exhibits the typical behavior of a Pygmy dipole state where the proton and
neutron
transition densities are in phase inside the nucleus, while at the surface region the contribution
comes from the neutron density only. Consequently, at the surface
the isoscalar and isovector transition densities have the same intensity giving rise to a strong isospin mixing.
For the calculated dipole states this behavior is supported by the present data, which manifests 
significant isospin mixing in the energy region under investigation.

\begin{figure}[b]
\vspace{1cm}
\includegraphics[scale=0.35]{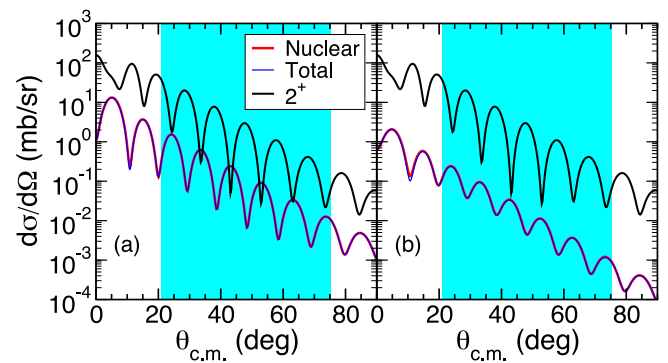}
\caption{\label{fig:Coulomb_effect_multipole} (Color online) Cross sections of inelastic scattering of $\alpha-$particles
are plotted as a function
of scattering angle in the center of mass frame for the 2$^+$ state in $^{74}$Ge at 596 keV and
dipole states at (a) 4.55 and (b) 7.05 MeV.
The blue shaded areas represent the angular coverage of scattered $\alpha-$particles in the present measurement.}
\end{figure}

An estimate of the inelastic cross section of states due to different reaction mechanisms is
obtained using the TALYS 1.6 reaction code \cite{talys}.
These calculations suggest that for a 48 MeV $\alpha$ beam, the compound reaction does not
contribute at any excitation energy under consideration, while for $E_x \approx$ 6 MeV the contribution
from pre-equilibrium reactions is an order of magnitude less than that from direct reactions and
gradually increases with $E_x$. Therefore, a direct comparison with other experimental data
should be taken with some degree of caution at the highest excitation energies.

In principle, the presence of the Coulomb interaction
between the target and projectile has the capability to 
substantially contribute to the observed cross-sections \cite{Lanza2014}. 
To investigate the effect of the Coulomb interaction on the observed inelastic cross sections, theoretical cross sections were calculated both
with and without taking the Coulomb interaction into account. These theoretical cross sections were obtained for the dipole states
at $E_x$ = 4.55 and 7.05 MeV by performing Distorted Wave Born Approximation (DWBA) calculations,
carried out using the FRESCO code \cite{fresco}.	
The radial nuclear form factors were constructed within a double folding procedure using the
microscopic transition densities of Fig.~\ref{fig:td},
see Ref.~\cite{Lanza2015} for more details on the procedure. For the Coulomb form factors we
have used the analytic expression built inside the FRESCO code.
For these calculations the double folding potential was used as the real part of the optical potential, while
for the imaginary part the same geometry as for the real part but with half the intensity was chosen \cite{Lanza2015}.
These results are shown in Fig.~\ref{fig:Coulomb_effect_multipole} where a negligible difference between the calculations
performed using only the nuclear interaction (red curve) and using both the nuclear and
Coulomb interaction (blue curve) is observed for the detection angles under study (blue shaded areas).
 For these low-lying dipole states it has been shown that the nuclear and Coulomb contributions interfere constructively
in the nuclear surface region \cite{Lanza2014}.
This feature is expected not to be visible for this relatively low
incident energy since the Coulomb contribution becomes important
as the beam energy increases towards 30 MeV/u \cite{Lanza2011, Lanza2014}.
We are aware of the fact that, while the relation between the inelastic cross section and the
$B_{em}(E1)$ is clear for the Coulomb excitation (they are
proportional), the relation between the isoscalar response and the
inelastic excitation cross section due to an isoscalar probe is not so evident.
In fact, the ratio between the $B_{is}(E1)$ of the two states at 4.55 and 7.05 MeV is 2.2 while
the ratio between the corresponding values of the cross
sections is 6.4 at the first maximum. If we eliminate the effect of the Q-value, by placing the two states at the same energy, then the ratio
decreases to 4.1, still far from 2.2. However, in Ref.~\cite{Lanza2014} a calculation of
the cross section was presented in the framework of a semiclassical
model, that provides the missing link to directly compare the results from the microscopic
RQTBA calculations to experimental data measured via
the ($\alpha$, $\alpha'$$\gamma$) reaction, confirming the structural splitting of
the low-lying $E1$ strength.

It is instructive to also have an estimate of the cross-section of states with higher multipolarities.
Therefore, we also performed calculations for the first-excited 2$^+$ state in $^{74}$Ge, using a
collective macroscopic nuclear form factor. The $B(E2)$ value of the 596 keV transition is taken
to be 3050 $e^2fm^4$ from Ref.~\cite{leconte1980} with a deformation length of 1.43 fm.
The results are shown in Fig.~\ref{fig:Coulomb_effect_multipole}, where the cross-sections for
the 2$^+$ state (black curve) are significantly higher when compared to the dipole states. This is
not only the case for the detection angles of the present
experiment, but also for very forward angles.

It is interesting to point to a recent measurement of the photon strength function
below the neutron separation energy in $^{74}$Ge \cite{renstrom2016}, using the so-called
Oslo Method. Despite the limited $\gamma$-ray detection resolution, a broad structure is
observed in the 6 $< E_\gamma <$ 8 MeV range. It is highly probable that this feature is the
same Pygmy dipole resonance structure as observed in this work.  

\section{SUMMARY AND CONCLUSION}

We provide new results, which indicate a suppression in relative cross section for
the excitation of the PDR in $^{74}$Ge populated through inelastic $\alpha-$scattering, 
when compared to photon scattering data for $E_x>6$ MeV. The observed dipole response splits into two distinct parts: one at lower energy,
with excitations that have strong isospin mixing, and one at higher energy with predominant isovector character. 
The results are particularly important in improving our understanding of the emergence and persistence of the PDR for
low $N/Z$ nuclei. As such, measurements in other mass regions are undoubtedly necessary to fully understand the
evolution of the PDR from near-isospin saturated systems towards nuclei with large $N/Z$ ratios. Finally, the
present work highlights the importance of using complementary probes to photon scattering, in order to reveal detailed information about the underlying nature of dipole excitations.

\section{Acknowledgments\protect\\}

The authors would like to thank the operational staff at iThemba LABS for providing excellent beam
quality throughout the experiment and Lawrence Berkeley National Laboratory for making available the
$^{74}$Ge target. This work was
supported by the National Research Foundation of South Africa under Grant No. 92789, and No. 93500; by the Research Council of Norway,
Project Grants No. 205528, No. 213442, and No. 210007; by US-NSF Grants No. PHY-1204486 and No. PHY-1404343;
by the US Department of Energy under Contracts No. DE-AC52-07NA27344, and No. DE-AC02-05CH11231; and by ERC-STG-2014
Grant No. 637686.


\begin{thebibliography}{60}
\bibitem{paar2007} N.~Paar, D.~Vretenar, E.~Khan, and G.~Col{\`o}, Rep. Prog. Phys. {\bf 70}, 691 (2007).
\bibitem{savran2013} D.~Savran, T.~Aumann, and A.~Zilges, Prog. Part. Nucl. Phys. {\bf 70}, 210 (2013).
\bibitem{bracco2015} A.~Bracco, F.~C.~L.~Crespi and E.~G.~Lanza, Eur. Phys. J. A {\bf 51}, 99 (2015).
\bibitem{lane1971} A.~M.~Lane, Ann. Phys. {\bf 63}, 171 (1971).
\bibitem{reinhard2013} P.~-G.~Reinhard and W.~Nazarewicz,  Phys. Rev. C {\bf 87}, 014324 (2013).
\bibitem{pellegri2014} L.~Pellegri \textit{et al.}, Phys. Lett. B {\bf 738}, 519 (2014).
\bibitem{crespi2015} F.~C.~L.~Crespi \textit{et al.},  Phys. Rev. C {\bf 91}, 024323 (2015).
\bibitem{poltoratska2012} I.~Poltoratska \textit{et al.}, Phys. Rev. C {\bf 85}, 041304(R) (2012).
\bibitem{krumbholz2015} A. M. Krumbholz \textit{et al.}, Phys. Lett. B {\bf 744}, 7 (2015).
\bibitem{savran2006} D.~Savran \textit{et al.}, Phys. Rev. Lett. {\bf 97}, 172502 (2006).
\bibitem{endres2009} J.~Endres, \textit{et al.}, Phys. Rev. C {\bf 80}, 034302 (2009).
\bibitem{poelhekken1992} T.~D.~Poelhekken \textit{et al.}, Phys. Lett. B {\bf 278}, 423 (1992).
\bibitem{endres2010} J.~Endres, \textit{et al.}, Phys. Rev. Lett. {\bf 105}, 212503 (2010).
\bibitem{derya2012} V.~Derya \textit{et al.}, J. Phy. Conf. Ser. {\bf 366}, 012012 (2012).
\bibitem{derya2013} V.~Derya \textit{et al.}, Nucl. Phys. {\bf A906}, 94 (2013).
\bibitem{derya2014} V.~Derya \textit{et al.}, Phys. Lett. B {\bf 730}, 288 (2014).
\bibitem{crespi2014} F.~C.~L.~Crespi \textit{et al.}, Phys. Rev. Lett. {\bf 113}, 012501 (2014).
\bibitem{volz2006} S.~Volz \textit{et al.}, Nucl. Phys. {\bf A779}, 1 (2006).
\bibitem{govaert1998} K.~Govaert \textit{et al.}, Phys. Rev. C {\bf 57}, 2229 (1998).
\bibitem{romig2013} C.~Romig \textit{et al.}, Phys. Rev. C {\bf 88}, 044331 (2013).
\bibitem{shizuma2008} T.~Shizuma \textit{et al.}, Phys. Rev. C {\bf 78}, 061303 (2008).
\bibitem{schw2007} R.~Schwengner \textit{et al.}, Phys. Rev. C {\bf 76}, 034321 (2007).
\bibitem{ben2009} N.~Benouaret \textit{et al.}, Phys. Rev. C {\bf 79}, 014303 (2009).
\bibitem{mak2010} A.~Makinaga \textit{et al.}, Phys. Rev. C {\bf 82}, 024314 (2010).
\bibitem{tsoneva2008} N.~Tsoneva, and H.~Lenske, Phys. Rev. C {\bf 77}, 024321 (2008).
\bibitem{paar2009} N.~Paar \textit{et al.}, Phys. Rev. Lett. {\bf 103}, 032502 (2009).
\bibitem{martini2011} M.~Martini, S.~P{\' e}ru, and M.~Dupuis, Phys. Rev. C {\bf 83}, 034309 (2011).
\bibitem{tohyama2012} M.~Tohyama, and T.~Nakatsukasa, Phys. Rev. C {\bf 85}, 031302(R) (2012).
\bibitem{vretenar2012} D.~Vretenar \textit{et al.}, Phys. Rev. C {\bf 85}, 044317 (2012).
\bibitem{nakada2013} H.~Nakada, T.~Inakura, and H.~Sawai, Phys. Rev. C {\bf 87}, 034302 (2013).
\bibitem{goriely1998} S.~Goriely, Phys. Lett. B {\bf 436}, 10 (1998).
\bibitem{goriely2004} S.~Goriely, E.~Khan, and M.~Samyn, Nucl. Phys. {\bf A739}, 331 (2004).
\bibitem{litvinova2009} E.~Litvinova \textit{et al.}, Nucl. Phys. {\bf A823}, 26, (2009).
\bibitem{daoutidis2012} I.~Daoutidis, and S.~Goriely, Phys. Rev. C {\bf 86}, 034328 (2012).
\bibitem{piekarewicz2006} J.~Piekarewicz, Phys. Rev. C {\bf 73}, 044325 (2006).
\bibitem{piekarewicz2011} J.~Piekarewicz, Phys. Rev. C {\bf 83}, 034319 (2011).
\bibitem{rosier1989} L.~Rosier, and E.~I.~Obiajunwa,  Nucl. Phys. A {\bf 500}, 323 (1989).
\bibitem{moeller2008} P.~M{\"o}ller \textit{et al.}, At. Data Nucl. Data Tables 94, 758 (2008).
\bibitem{jung1995} A.~Jung \textit{et al.},  Nucl. Phys. {\bf A584}, 103 (1995).
\bibitem{massarczyk2015} R.~Massarczyk \textit{et al.}, Phys. Rev. C {\bf 92}, 044309 (2015).
\bibitem{afrodite} M.~Lipoglav{\v s}ek \textit{et al.}, Nucl. Instr. Meth. Phys. Res. A {\bf 557}, 523 (2007).
\bibitem{micron} Micron Semiconductor, Product Cataloque, http://www.micronsemiconductor.co.uk/pdf/w1.pdf
\bibitem{XIA} http://www.xia.com
\bibitem{root} https://root.cern.ch
\bibitem{eidson1962} W.~W.~Eidson and J.~G.~Cramer, Jr., Phys. Rev. Lett. {\bf 9}, 497 (1962).
\bibitem{schurmann1987} B.~Sch{\" u}rmann \textit{et al.},  Nucl. Phys. {\bf A475}, 361 (1987).
\bibitem{litvinova2008} E.~Litvinova, P.~Ring, and V.~Tselyaev, Phys. Rev. C {\bf 78}, 014312 (2008).
\bibitem{ring1996} P.~Ring, Prog. Part. Nucl. Phys. {\bf 37}, 193 (1996).
\bibitem{vretenar2005} D.~Vretenar \textit{et al.}, Phys. Rep. {\bf 409},  101  (2005).
\bibitem{litvinova2010} E.~Litvinova, P.~Ring, and V.~Tselyaev, Phys. Rev. Lett. {\bf 105}, 022502 (2010).
\bibitem{litvinova2013} E.~Litvinova, P.~Ring, and V.~Tselyaev, Phys. Rev. C {\bf 88}, 044320 (2013).
\bibitem{lalazissis2009} G.~A.~Lalazissis \textit{et al.}, Phys. Lett. B {\bf 671}, 36 (2009).
\bibitem{nndc} http://www.nndc.bnl.gov (as of April 2015).
\bibitem{talys} A.~J.~Koning \textit{et al.}, \emph{Nuclear Data for Science and Technology}, edited by O. Bersillon \textit{et al.} (EDP Sciences, Nice, France, 2008), p. 211 (see also http://www.talys.eu).      
\bibitem{Lanza2014} E.~G.~Lanza \textit{et al.}, Phys. Rev. C {\bf 89}, 041601(R) (2014).
\bibitem{fresco} I.~J.~Thompson, Comp. Phys. Rep. {\bf 7}, 167 (1988); http://www.fresco.org.uk
\bibitem{Lanza2015} E.~G.~Lanza, A.~Vitturi, and M.~V.~Andr{\'e}s, Phys. Rev. C {\bf 91}, 054607 (2015).
\bibitem{Lanza2011} E.~G.~Lanza \textit{et al.}, Phys. Rev. C {\bf 84}, 064602 (2011).
\bibitem{leconte1980} R.~Leconte \textit{et al.}, Phys. Rev. C {\bf 22}, 2420 (1980).
\bibitem{renstrom2016} T.~Renstr{\o}m \textit{et al.}, Phys. Rev. C {\bf 93}, 064302 (2016).


\end{thebibliography}
\end{document}